# Observation of photonic 'Fermi arcs' in topological metamaterials

Biao Yang[1†], Qinghua Guo[1, 3†], Ben Tremain[2†], Lauren E. Barr[2], Wenlong Gao[1], Hongchao Liu[1], Benjamin Béri[1], Yuanjiang Xiang[3], Dianyuan Fan[3], Alastair P. Hibbins[2]*, Shuang Zhang[1]*

[1] School of Physics and Astronomy, University of Birmingham, Birmingham B15 2TT, United Kingdom.

[2] Electromagnetic and Acoustic Materials Group, Department of Physics and Astronomy, University of Exeter, Stocker Road, Exeter, EX4 4QL, United Kingdom.

[3] Key Laboratory of Optoelectronic Devices and Systems of Ministry of Education and Guangdong Province, College of Optoelectronic Engineering, Shenzhen University, Shenzhen 518060, China.

†These authors contributed equally to this work.

*Correspondence to: A.P.Hibbins@exeter.ac.uk; s.zhang@bham.ac.uk

**The discovery of topological phases introduces new perspectives and platforms for various interesting physics originally investigated in quantum context[1-5] and then, on an equal footing, in classic wave systems, such as photonics[6-13], acoustics[14-16] and mechanics[17]. As a characteristic feature, nontrivial Fermi arcs, connecting between topologically distinct Fermi surfaces, play vital roles in the clarification of Dirac and Weyl semimetals[18-20] and have been observed in quantum materials very recently[21-23]. However, in classical systems, no direct experimental observation of Fermi arcs in the momentum space has been reported so far. Here, using near-field scanning measurements, we show the observation of a photonic 'Fermi arcs' connecting topologically distinct bulk states in chiral hyperbolic metamaterials. To verify the topological nature of this system, we further observe backscattering-**

immune propagation of nontrivial surface wave across a three-dimension physical step. Our results demonstrate metamaterial approach towards topological photonics and offer a deeper understanding of topological phases in three-dimension classical systems.**

In order to observe the characteristics of a Weyl semimetal, a sample that possesses either broken time-reversal symmetry or broken inversion symmetry is required. In classical electromagnetism, it is more feasible to break inversion symmetry than time-reversal symmetry since breaking time-reversal symmetry requires lossy magnetic materials and external magnetic fields. Here a chiral hyperbolic metamaterial (CHM) with broken inversion symmetry, which was proposed as topological metamaterial[24], is employed to explore the photonic 'Fermi arcs'. The topological nature of the metamaterial can be described by a homogeneous effective model and its Weyl points arise from the degeneracies between intrinsic electromagnetic modes: longitudinal plasmonic mode and spin-polarized transverse mode[24,25]. It is different from usual photonic realization of Weyl degeneracy in photonic crystal[26,27], where spatial degrees of freedoms offer the state sub-space. The two bands that cross each other forming the Weyl cone have the same sign of velocity along a certain momentum space direction ($k_y$ as shown in Fig. 1a and b). In a simple effective model (see the supplementary information) of the CHM, one pair of type-II Weyl points[24,25], represented by the two red spots (W1s) in Fig. 1c, is located at large k therefore becomes much easier to be identified experimentally. Without any mirror symmetries, two blue spots W2s (one is hidden), located at a lower frequency in Fig. 1c are chiral partners of the two W1 points. Both W1 (W2) Weyl nodes are related to each other by time reversal symmetry and have the same chirality. Nontrivial surface state between the Weyl point partners (W1 and W2) is also shown in Fig. 1c (red surface), which serve as a signature of the topological nature of the system. Band structure along high symmetry lines is showed in Fig. 1d. The nontrivial gap where the topological surface states reside is also indicated. In Fig. 1e surface state arcs connecting positive and negative Weyl points with varying frequency are projected onto $k_x$-$k_y$ plane. Its three-dimension view can be found in Fig. 1c as the tilted cyan line. In order to demonstrate the evolution of surface state with frequency, three Equi-Frequency Contours (EFCs) are showed in Fig. 1f, g and h. In the Fig.1f, clearly one sees surface state arcs abruptly arise from W2 Weyl points, whose bands along ΓX are flat (dashed line), behaving like a transition phase between type-I and type-II Weyl points[28]. As the frequency increasing, surface state arcs will be tangent to the corresponding bulk states, which contain the projections of Weyl points. Finally, they terminate at W1s in Fig. 1h.

The CHM is constructed by stacking of two-dimensional tri-layer unit. Fig. 2a illustrates a cubic unit cell of the CHM, which is fabricated on copper-clad FR4 substrates with dielectric constant of 4.1. The thickness of each copper layer is 35 μm and the copper can be regarded as a perfect electric conductor (PEC) in the studied frequency range 5-10 GHz. To obtain the desired hyperbolic properties of the CHM, 200 μm-wide metallic wires are formed along the *y*-direction on top surface of the bottom layer (Fig. 2a). Metallic crosses are superimposed on these wires to increase the local capacitances and suppress the strong non-local effects induced by the metallic wires alone[29]. In order to break inversion symmetry, we introduce metallic helix structure on the top layer. Due to the small number of turns per unit length, electric current driven along the helix induces a magnetic dipole moment that is slightly misaligned with the *x*-axis[30]. This leads to a tiny shift ($0.01\pi/a$) of W1 away from the $k_y$ axis and a large shift ($0.14\pi/a$) of W2 away from $k_x$ axis as shown in Fig 2d. Nevertheless, they are still located on the $k_z = 0$ plane due to both the time-reversal symmetry and $C_2$ rotational symmetry along the *z*-axis. The large momentum separation of the Weyl points, as shown in Fig. 2d, allows us to easily resolve topological 'Fermi arcs' from the Fourier transformation of near field scanning. Upon the realistic structure,

Fig. 2e shows simulated band crossing in the vicinity of the type-II Weyl node (W1). The simulated dispersion around W2 is presented in the supplementary information.

The near field scans are first conducted on both top (*xy*) and side (*yz*) surfaces of CHM with the same polarized source and probe, i.e. Z (polarized source) - Z (polarized probe) for top, and X (polarized source) - X (polarized probe) for side. After Fourier transformation of the near field spatial distribution, we obtain Fig. 3a and b presenting the 'Fermi surfaces' (both bulk and surface states, i.e., EFCs) at 5.46 GHz for the top and side surfaces, respectively. As can be seen, 'Fermi arcs' are tangent and terminated to their corresponding bulk modes in both top and side cases. Fig. 3c and d show the top- and side-scanned Fermi surfaces at 8.13 GHz respectively. From Fig. 3c, one can see a resemblance to the modelled density of states (DOS) in Fig. 1a, a feature that provides verification of type-II Weyl dispersion. As previously discussed, the misalignment of the helix axis with the underlying lattice results in a shift of the W2 Weyl points away from $k_y$ axis. However, this shift is too small to be observed in our experiment due to limited k-space resolution with minimum pixel of $0.027\pi/a$. In Fig. 3d, a similar DOS 'crossing' can be seen from the side surface scan, as schematically showed in Fig. 1b. It is worth pointing out that W2 Weyl points (5.15 GHz) appear accompanying with strong resonance from metallic helices and cannot be clearly recognized in experiment. These measured bulk/surface band structures are consistent with the simulations (from CST Microwave Studio) in terms of the 'Fermi surfaces' and their locations in the momentum space, as shown in Fig. 3. The deviations between the measurements and calculations regarding the exact shape and size of each 'Fermi surface' may arise from the sample fabrication errors and misalignment of layer stacking. Extra comparisons between experiment and simulation results at different frequencies are given in supplementary information.

In general, the scanned near field intensity profile can vary significantly with the measurement configuration, such as the proximity to the sample surface and the polarizations of source and probe, which offers great potential for topological near field sensing. Our near field scanning measurements on the top and side surface configurations with different polarization combinations can provide complementary information for revealing the topological features of CHM. Fig. 4a and b show a surface wave on the top surface with X (polarized source) - Z (polarized probe) configuration. Both real space and momentum spaces show the strong excitation of surface state in the nontrivial gap by the antenna source. The intensity contribution from bulk bands, on the other hand, is largely suppressed and decay rapidly away from the source position ($x = 150$mm, $y = 0$mm). When the surface wave impinges on a surface edge which is invariant along $y$ direction, the topologically nontrivial surface wave is expected to travel around the right-angle edge and continue propagating on the side surface ($y$-$z$ plane). Fig. 4d shows the results when an X-polarized probe scans the side surface with the source placed at top surface ($x = 10$mm, $y = 0$mm). One can observe that the bulk modes are dramatically weakened by edge backscattering but the surface wave is not, which can also be confirmed by its bright topological 'Fermi arcs' as shown in Fig. 4d compared with Fig. 3b.

In order to demonstrate the topological protection of the surface state, a step configuration is created by stacking a metamaterial block on top of another with a 104 mm shift in the $x$ direction, as shown in Fig. 4e. A surface wave is launched at the upper surface, and the probe scans the normal field distribution in the upper, side and lower surfaces at frequency of 5.46 GHz. It is observed that the surface wave conformally bends around the step and keeps propagating forward without being reflected by the edges. It should be noted that since our probe antenna cannot reach the lower corner (built by side and lower surfaces) due to thick cladding layer of the antenna cable, there exists a discontinuity in the measured phase across the lower corner. This measurement of robust propagation of surface wave across the step serves as a direct observation of the topological protection of 'Fermi arcs'.

In conclusion, we have made the direct observation of photonic 'Fermi arcs' at the surfaces of a photonic type II Weyl metamaterial. Our work may provide new insights and open many new avenues to the surface photonics. Due to the robustness of topologically protected interfacial wave propagation and other practical advantages in photonic realm, such as artificially controllable structure design, we anticipate our observation to be a start point for surface 'periscope' imaging technology, near field sensing and directional information transmission in bulky integrated photonics.

**Methods**

The experimental setup is schematically shown in Fig. 2c (see supplementary information for real setup). Instead of using angle-resolved transmission measurement, we employ a microwave vector network analyser (VNA) and a near-field antenna acting as source (stationary) to provide excitation of electromagnetic surface waves, which are subsequently probed with a second near-field antenna (controlled by a *xyz* translation stage). Both the amplitude and phase of the electric field near the surface of the CHM are measured. The band structure of surface wave can then be determined from Fourier analysis of the spatial distribution of electric field at each frequency. In the scanning measurements, the scan step is set at 2 mm, which equalling to half lattice constant of the cubic unit cell ($4\times4\times4$ mm$^3$) will determine the maximum surface k-space range as $[-2\pi/a, 2\pi/a]^2$ under Fourier transformation. With fixed scan step, the k space resolution is controlled by the maximum area being scanned, i.e., the size of sample fabricated. Here each unit cell can be considered as an electromagnetic 'meta-atom' with structure-period/ vacuum-wavelength ratio around 1/10. The collective effect of a large number of these 'meta-atoms' is measured, which is the key principle of metamaterials in general. Because the type-II Weyl point is located far away from the light cone, the topological surface wave is well confined at the interface between the CHM and air, which greatly facilitates the near-field measurements.

**Acknowledgments** We thank Ling Lu for discussions and feedback. This work was financially supported by ERC Consolidator Grant (Topological). B. Y. acknowledges China Scholarship Council (201306110041). S. Z. and B. B. acknowledge support from the Royal Society. Y. X. and D. F. acknowledge support from the National Natural Science Foundation of China (Grant No. 61490713) and the Science and Technology Planning Project of Guangdong Province (Grant No. 2016B050501005). L. E. B. and A. P. H. acknowledge financial support from EPSRC of the United Kingdom (Grant No. EP/L015331/1).

**Author Information** The authors declare no competing financial interests. Correspondence and requests for materials should be addressed to A.P.H. (A.P.Hibbins@exeter.ac.uk) and S.Z. (s.zhang@bham.ac.uk).

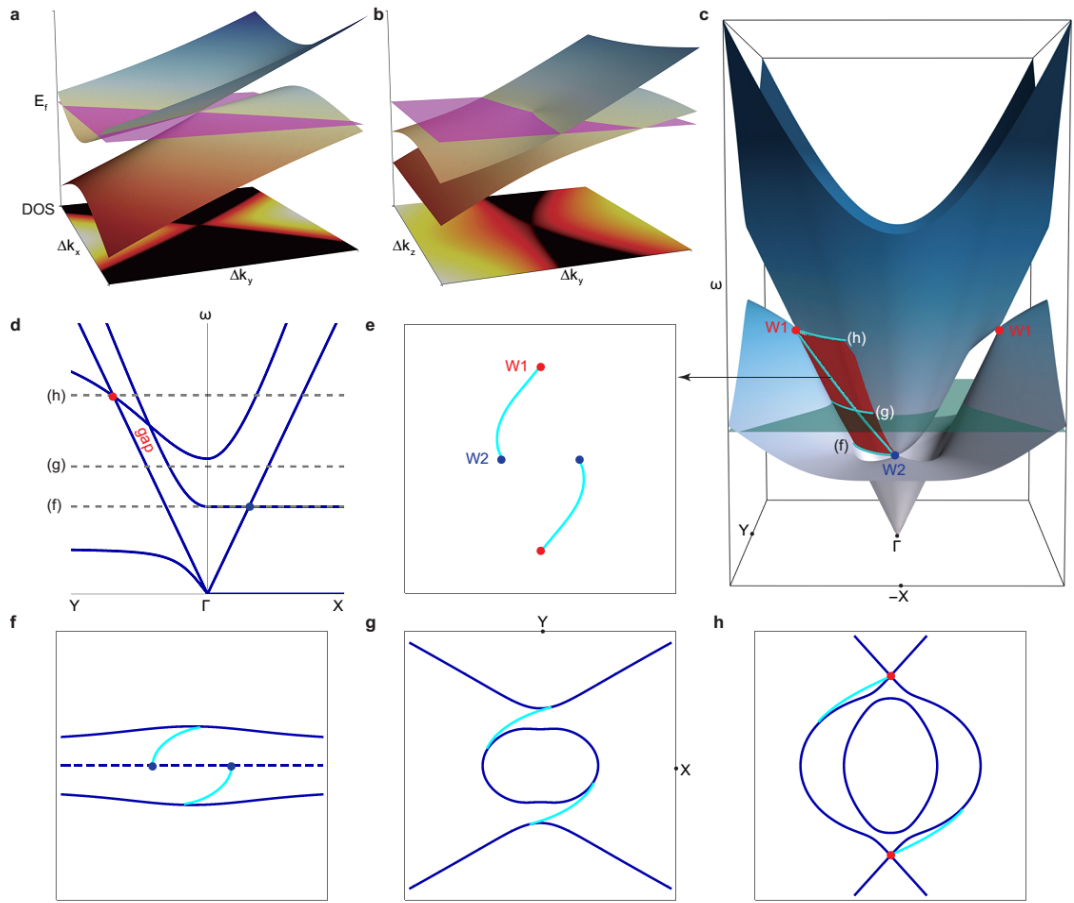

**Figure 1| Bulk and surface states in effective media model of chiral hyperbolic metamaterials (CHM).** Schematic view of energy dispersion near Type-II Weyl point for CHM with respect to **(a)** *x-y* and **(b)** *y-z* momentum space. Bottom planes show momentum resolved Density of States (DOS) at the 'Fermi energy' $E_F$, and $E_F$ is indicated by a purple plane. **c,** Effectively modelled band structures of bulk (7$^{th}$ and 8$^{th}$ bands) and surface states for the CHM. Red spots W1s indicate one pair of type-II Weyl points. Blue spots W2s (one is hidden) are the chiral partners of the red ones W1s. One surface state between two topological partners is indicated by the red surface, upon which the cyan lines highlight surface state arcs. In order to show the gap, the other surface state is not plotted, which can be obtained after time reversal operation of the present one. **d**, Bulk bands along Y(0,-3,0)-Γ(0,0,0)-X(3,0,0). **e**, Surface band structure on a varying-frequency $k_x$-$k_y$ map. The arc ($k_y$>0) is constructed from intersection

between the surface state (red surface in (c)) and a plane with constraints: ω is proportional to $k_y$ and $k_x$ is arbitrary. **f**, **g** and **h**, EFCs of bulk and surface states on $k_x$-$k_y$ plane. Three different frequencies are showed as indicated in (c) and (d). Here, $\varepsilon_0=\mu_0=c=1$, ω, $k_x$ and $k_y$ are normalized with respect of $\omega_p$ (at $k_y=0$), which is the plasma frequency.

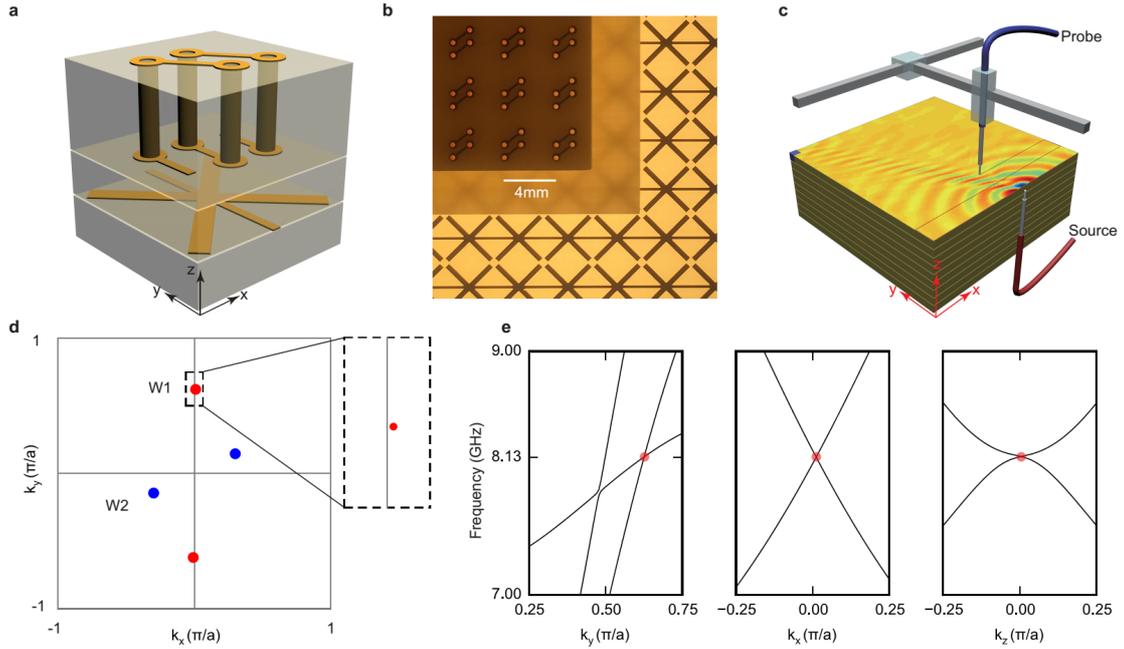

**Figure 2| Photonic type-II Weyl metamaterial realized within a CHM. a**, Cubic unit cell of CHM with side length a=4 mm, consisting of three layers, bottom hyperbolic layer (1mm), middle blank layer (1mm) and top chiral layer (2mm). Each helix has 2.5 turns, with its axis along the *x*-direction. Its length and cross-sectional area are optimized to give a fundamental resonance frequency around 5 GHz. The blank layer (FR4) between the chiral and hyperbolic layer is designed to avoid shorting contact between them. **b**, Tri-layer sample fabricated with printed circuit board technology. There are 75 unit cells (4 × 4 mm$^2$) along the in-plane directions on each layer. On the hyperbolic layer, metallic lines go through the whole layer along y direction. **c**, Experimental setup and layer-stacking geometry. The field pattern represents the real experimental data scanned under *X* (polarized source) – *Z* (polarized probe)

configuration at 5.82 GHz. Surface wave is excited by one antenna source, another antenna serves as probe scanning the propagating near field. **d**, The slice of First Brillouin Zone with respect of $k_z=0$ and locations of Weyl points for W1: $(\pm0.01,\pm0.62)\pi/a$ and W2: $(\pm0.30,\pm0.14)\pi/a$. The zoomed-in rectangle indicates that W1 locates slightly away from $k_y$ axis. **e**, Simulated linear degeneracy around type-II Weyl point W1 (8.13 GHz) with realistic structure designed in CST microwave studio.

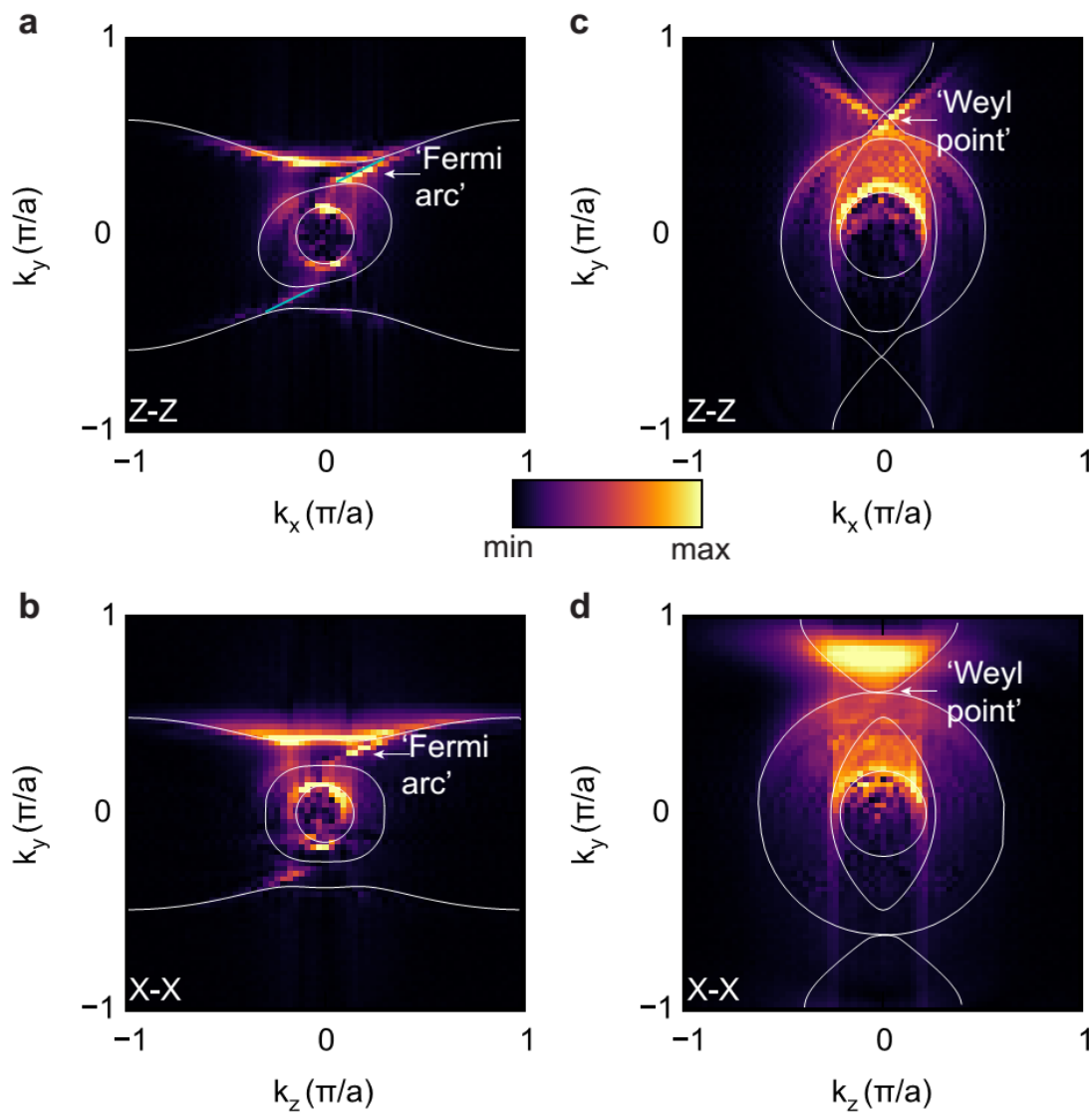

**Figure 3| Photonic 'Fermi surfaces' at two different frequencies.** Scanned with the configuration of *Z* (polarized source) - *Z* (polarized probe) on the top surface at (**a**) 5.46 GHz and (**c**) 8.13 GHz. Scanned with the configuration of *X* (polarized source) - *X*

(polarized probe) on the side surface at (**b**) 5.46 GHz and (**d**) 8.13 GHz. For X polarized source, the antenna orients –x direction. Middle solid circles indicate air-light cone. The other solid curves present simulation results from microwave CST studio. Data in $k_y < 0$ range is back-scattered from *y*-maximum boundary. 'Fermi arcs' and type II Weyl points are indicated by the white arrows. The crosses in (**c**) and (**d**) show another characteristic feature of type II Weyl semimetal as showed in Fig. 1(**a**) and (**b**) DOS planes.

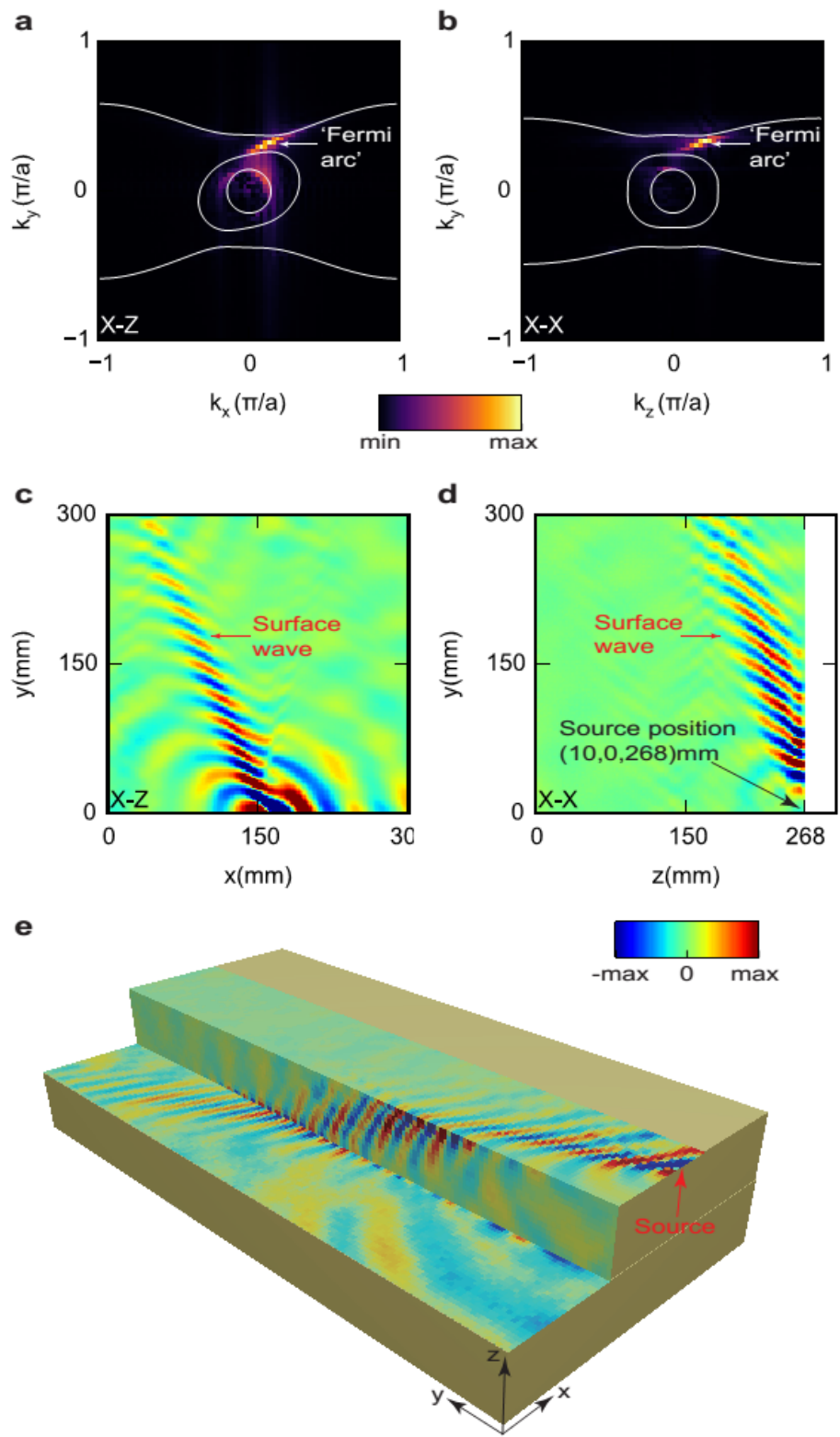

**Figure 4| Topologically protected surface waves and corresponding 'Fermi arcs'.**
Top surface scan taken with the *X* (polarized source) - *Z* (polarized probe) configuration at 5.46 GHz. **a**, Real-space instantaneous field. **c**, Momentum space amplitude field, being Fourier transformed from **(a)**. Side surface scan with respect of $X$ (polarized source) - $X$ (polarized probe) configuration at 5.46 GHz. **(b)** and **(d)** present the same plot as **(a)** and **(c)** but on the side surface. The excitation source position for the side surface scan is positioned on top surface with coordinate indicated. Solid circles indicate air-light cone, which are far away from 'Fermi arcs', thus surface wave is being well localized. **e**, Backscattering-immune surface wave propagating on 3D 'step' experimental geometry. Source is set with $X$-polarization, while normal component of electric field is probed for all surfaces. The step width, height, length are 104 mm (both for upper and lower surfaces), 60 mm and 600 mm, respectively. (See real experimental setup in supplementary information).

14